\documentclass[aps,prl,twocolumn,superscriptaddress,floatfix,nofootinbib]{revtex4-2}

\usepackage{graphicx}
\usepackage{amsmath,amssymb}
\usepackage{bm}
\usepackage{dcolumn}
\usepackage{xcolor}

\newcommand{\rev}[1]{#1}
\newcommand{\revcolor}{}

\begin{document}

\title{Stable Ballistic Prograde Cyclers in the Three-Body Problem}

\author{Shane D. Ross}
\email{sdross@vt.edu}
\affiliation{Aerospace and Ocean Engineering, Virginia Tech, Blacksburg, Virginia 24061, USA}

\author{Michael Roberts-Tsoukkas}
\affiliation{Aerospace and Ocean Engineering, Virginia Tech, Blacksburg, Virginia 24061, USA}

\date{\today}

\begin{abstract}
We report the first continuous families of stable, ballistic, prograde cycler orbits in the circular restricted three-body problem: periodic trajectories that alternately undergo temporary capture and orbit each primary.
We construct continuous families of symmetric cyclers from
intersections of the stable and unstable manifold tubes of the
$L_1$ Lyapunov orbit and exhibit stable examples across more than
two orders of magnitude in mass ratio, from the Sun--Jupiter regime
to the equal-mass limit.
Linear stability separates naturally into planar and
out-of-plane components.
A planar-stable branch of every computed family is created
simultaneously with
a hyperbolic branch in a saddle-center bifurcation of the return map
at the family's maximal Jacobi constant, while out-of-plane
instability occurs only through isolated parametric resonances.
Every computed family contains a subfamily that is linearly stable to both planar and out-of-plane perturbations.
These stable cyclers persist into the full three-body problem
for sufficiently small third mass.
We conjecture that saddle-center birth is universal
among cycler families, suggesting that stable cyclers are a generic
feature of three-body dynamics.
\end{abstract}

\maketitle

The periodic solutions of the three-body problem remain, in Poincar\'e's view, the principal opening through which an otherwise intractable dynamics can be understood~\cite{Poincare1892}. 
Their numerical and analytic discovery continues to be fruitful, from the figure-eight orbit~\cite{Moore1993,ChencinerMontgomery2000,DoPaKeDiGaVa2003} to the topological families of Newtonian three-body orbits found more recently~\cite{SuvakovDmitrasinovic2013}. 
In the restricted problem, a distinguished class are the \emph{cyclers}: 
periodic trajectories that alternately undergo temporary capture and complete one or more prograde revolutions about each primary.
They are important as natural transport pathways in cislunar space and
as candidate periodic configurations in binary gravitational
systems.

Cyclers have been studied for decades. 
Newton considered periodic trajectories passing repeatedly near two gravitating masses~\cite{rN59}.
Arenstorf later proved the existence of periodic trajectories with retrograde encounters of the secondary~\cite{arenstorf1963existence}.
Davidson computed the earliest numerical examples of the prograde class studied here~\cite{Davidson1964}, which Szebehely presumed unstable~\cite{Szebehely1967}. 
Aldrin later proposed Earth--Moon cyclers for routine logistics, albeit with periodic propulsion~\cite{Aldrin1985}.
Subsequent families~\cite{uphoff1993lunar,casoliva2008families,casoliva2010two,GenovaAldrin2015,de2022low,wilmera2022earth,wittal2022earth} are, to our knowledge, uniformly retrograde, propulsive, or linearly unstable. 
Leiva and Briozzo reported unstable periodic orbits
related to the prograde class studied here~\cite{leiva2006control,
leiva2008extension}.
Whether stable ballistic prograde cyclers exist has  remained an open question.
 
\begin{figure}[t]
\centering
\includegraphics[width=\columnwidth]{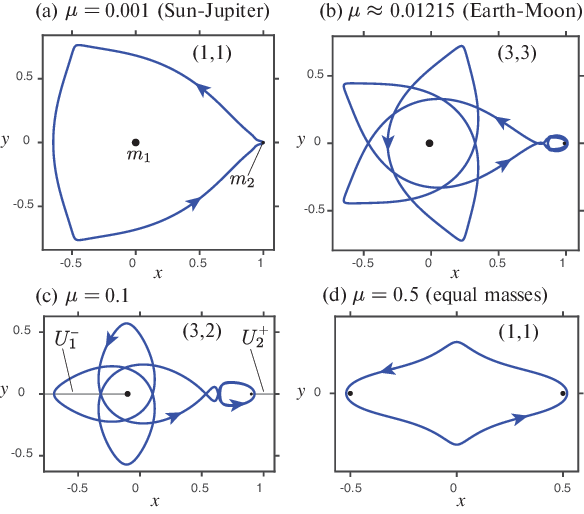}
\caption{
Stable, ballistic, prograde cyclers in the rotating frame. 
The primaries are shown as black markers, and arrows indicate the direction of motion.  
(a) Sun--Jupiter regime, $\mu=0.001$, $(1,1)$. 
(b) Earth--Moon, $\mu\approx0.01215$, $(3,3)$. 
(c) $\mu=0.1$, $(3,2)$. 
(d) Equal masses, $\mu=0.5$, $(1,1)$. 
Panel (c) indicates the Poincar\'e sections $U_1^-$ and $U_2^+$; their crossings define the indices $(k_1,k_2)$.
}
\label{fig:gallery}
\end{figure}

Here we resolve this question by reporting the first continuous families of stable ballistic prograde cyclers.
We develop a geometric construction that generates continuous families of cyclers from intersections of invariant-manifold tubes associated with ballistic  capture~\cite{JaRoLoMaFaUz2002,DeJuLoMaPaPrRoTh2005}. 
Stable families exist across more than two orders
of magnitude in mass ratio, from the Sun--Jupiter regime
($\mu\approx0.001$) to the equal-mass limit
(Fig.~\ref{fig:gallery}).
Each computed family is born in a saddle-center bifurcation that creates stable and unstable branches simultaneously. 
We conjecture that this mechanism is universal among
cycler families; every computed family contains a subfamily that is
stable to both planar and out-of-plane perturbations.

\textit{Model.} 
The planar circular restricted three-body problem describes a massless particle moving in the gravitational field of two primaries of masses $m_1=1-\mu$ and $m_2=\mu$, fixed at $-\mu$ and $1-\mu$ on the $x$-axis in a uniformly rotating nondimensional frame, where $\mu \in (0,\tfrac12]$~\cite{Szebehely1967,KoLoMaRo2022}. 
With $r_1$ and $r_2$ the distances to the primaries, the equations of motion are,
\begin{equation}
\ddot{x}-2\dot{y} = -\bar U_x,   \quad
\ddot{y}+2\dot{x} = -\bar U_y,
\label{eq:eom}
\end{equation}
with the effective potential $\bar U=-\tfrac{1}{2}(x^2+y^2)-(1-\mu)/r_1-\mu/r_2$, and notation $\bar U_a = \partial \bar U/\partial a$. 
The  Jacobi integral,
\begin{equation}
\mathcal{C}=-2\bar U-(\dot{x}^2+\dot{y}^2),
\label{eq:jacobi}
\end{equation}
is conserved; motion at fixed $\mathcal C=C$ lies on the
three-dimensional energy manifold $\mathcal M_C$.
Cyclers require Jacobi constants $C<C_1$, where $C_1$ denotes the
Jacobi constant of $L_1$. At these values, transport between the
neighborhoods of the primaries becomes possible.

A $(k_1,k_2)$-cycler alternately completes $k_1$ and $k_2$ prograde circuits about $m_1$ and  $m_2$, respectively, \rev{per period, as counted in the rotating frame}.
We identify and classify cyclers by recording 
crossings of four Poincar\'e sections{\revcolor, defined on $\mathcal M_C$ by,
\begin{equation}
\begin{aligned}
U_1^\pm &= \{(x,\dot x)\,|\,y=0,\ \pm(x+\mu)>0,\ \pm\dot y>0\},\\
U_2^\pm &= \{(x,\dot x)\,|\,y=0,\ \pm(x-1+\mu)>0,\ \pm\dot y>0\},
\end{aligned}
\label{eq:sections}
\end{equation}
with $U_1^\pm$ in the $m_1$ realm, $U_2^\pm$ in the $m_2$ realm, and $\dot y$ obtained from $C$ through Eq.~\eqref{eq:jacobi} (Fig.~\ref{fig:gallery}(c)); we write $U_i=U_i^-\cup U_i^+$~\cite{RoRo2025Cyclers}}.
For prograde motion, $k_1$ and $k_2$ equal the numbers of
$U_1^-$ and $U_2^+$ crossings per period.

\textit{Construction.} 
For each $C<C_1$, a unique planar Lyapunov orbit about $L_1$ has two-dimensional stable and unstable manifolds $W^s$ and $W^u$ of cylindrical (tube) geometry~\cite{Conley1968,LlMaSi1985}.
Each tube has a branch into the neighborhood of each primary and forms a codimension-one separatrix in $\mathcal M_C$: trajectories inside a tube transit through the $L_1$ neck, whereas those outside do not~\cite{Conley1968,LlMaSi1985,KoLoMaRo2000,GoKoLoMaMaRo2004}.
Because cyclers repeatedly transit between the primaries,
they must lie in the intersection of the stable and unstable transport tubes. 
This geometric characterization reduces the search for cyclers to intersections of stable and unstable manifold tubes.

\begin{figure*}[t]
\centering
\includegraphics[width=\textwidth]{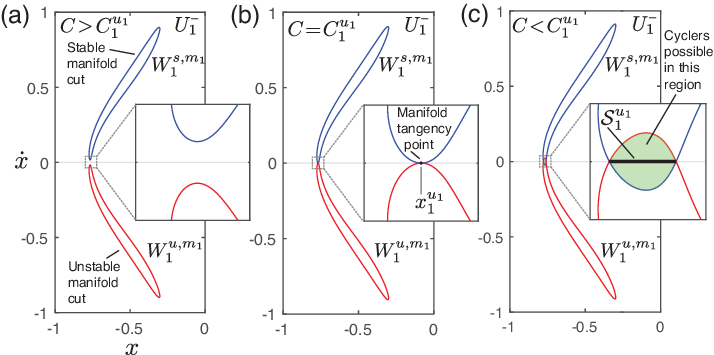}
\caption{
Geometric construction of symmetric cyclers on the Poincar\'e section $U_1^-$.
First cuts of the $m_1$ branches of the stable ($W^s$) and unstable ($W^u$) manifold tubes of the $L_1$ Lyapunov orbit are related by symmetry about ${\dot x=0}$.
Symmetric cyclers can occur only where the interiors of these cuts overlap; the symmetric initial conditions form the set $\mathcal S_1^{u_1}$ [Eq.~\eqref{eq:S}].
(a) For $C>C_1^{u_1}$ the cuts are disjoint and no symmetric cyclers exist. 
(b) At $C=C_1^{u_1}$ the cuts become tangent at $x_1^{u_1}$, marking the onset of overlap. 
(c) For $C<C_1^{u_1}$ the overlap region (shaded) contains the initial conditions of symmetric $(1,k_2)$ cyclers, while $\mathcal S_1^{u_1}$ is the corresponding symmetric subset along ${\dot x=0}$.
Shown for $\mu \approx 0.01215$ (Earth-Moon); the tangency-to-overlap geometry is the same for any $\mu \in (0, \tfrac{1}{2}]$.
}
\label{fig:construction}
\end{figure*}

Restricting attention to symmetric cyclers, whose initial conditions satisfy $\dot x=0$, 
we define,
\begin{equation}
\mathcal{S}_{k_1}^{u_1}\equiv U_1\cap\{\dot x=0\}\cap\mathrm{int}\,W^{s,m_1}_n\cap\mathrm{int}\,W^{u,m_1}_n,
\label{eq:S}
\end{equation}
where $W^{s,m_1}_n$ and $W^{u,m_1}_n$ denote the $n$th intersections with the
corresponding $U_1^\pm$ section; thus $k_1=2n-1$ on $U_1^-$ and $k_1=2n$ on
$U_1^+$.
An analogous set $\mathcal{S}_{k_2}^{u_2}$ is defined on $U_2$. 
For small $\Delta C=C_1-C$, the tube cross sections grow
proportionally to $\Delta C$~\cite{MacKay1990,ross2018experimental}.
At a critical value $C_{k_1}^{u_1}$,
the relevant stable and unstable cuts first become tangent along $\{\dot x=0\}$, 
marking the onset of the geometric overlap required for symmetric 
$(k_1,\cdot)$-cyclers (Fig.~\ref{fig:construction});
$C_{k_2}^{u_2}$ is defined analogously on $U_2$.
The largest Jacobi constant admitting a symmetric $(k_1,k_2)$-cycler is bounded by,
\begin{equation}
C_{(k_1,k_2)}=\min\bigl\{C_{k_1}^{u_1},\,C_{k_2}^{u_2}\bigr\}.
\label{eq:Cbound}
\end{equation}

\begin{figure}[!t]
\centering
\includegraphics[width=\columnwidth]{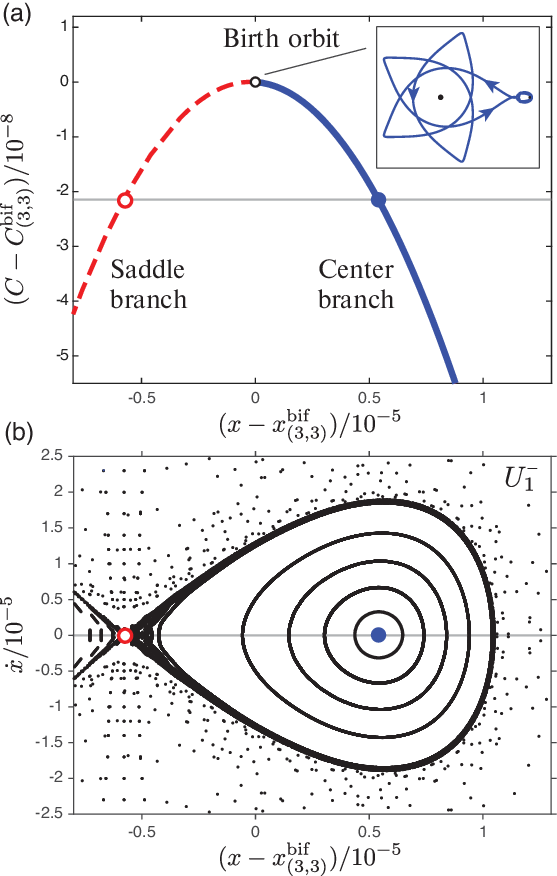}
\caption{
Saddle-center bifurcation creating the Earth–Moon $(3,3)$-cycler family. 
(a) Family near birth in the $(x,C)$ plane. The stable (solid) and unstable (dashed) branches meet at the saddle-center bifurcation point (open circle) at the family's maximal Jacobi constant, $C^{\rm bif}_{(3,3)}$. At the Jacobi constant indicated by the horizontal line, the family contributes one elliptic cycler (filled circle) and one hyperbolic cycler (open circle). Inset: the birth $(3,3)$-cycler in the rotating frame; at this scale, the cyclers at the selected Jacobi constant are visually indistinguishable from the birth orbit. 
(b) Return map on $U_1^-$ at the Jacobi constant indicated in (a). The elliptic fixed point (filled) is surrounded by invariant tori, while the hyperbolic fixed point (open) forms the companion member of the saddle-center pair.
}
\label{fig:bifurcation}
\end{figure}

For $C<C_{(k_1,k_2)}$,
candidate cyclers are obtained from the discrete intersection,
\begin{equation}
\Gamma = P(\mathcal{S}_{k_1}^{u_1})\cap\mathcal{S}_{k_2}^{u_2},
\label{eq:gamma}
\end{equation}
where $P$ is the section-to-section map.
Each point of $\Gamma$ defines a symmetric $(k_1,k_2)$-cycler.
The seeds are refined numerically by differential correction at fixed
$C$, exploiting the reversibility symmetry, $s_x:(x,y,\dot x,\dot y,t)\mapsto(x,-y,-\dot x,\dot y,-t)$, 
reducing the periodicity condition to a perpendicular crossing $\dot x=0$ at half period. 
Pseudo-arclength continuation~\cite{zhong2021differential} traces the resulting family in the $(x_0,C)$ plane (Fig.\ \ref{fig:bifurcation}, where $x_0 \in U_1$, while monitoring linear stability \cite{RoRo2025Cyclers}.
Each family is born at a maximum Jacobi constant $C_{(k_1,k_2)}^{\rm bif}<C_{(k_1,k_2)}$ 
at a corresponding point $x_{(k_1,k_2)}^{\rm bif}$.

\textit{Stability.} 
To assess the stability of planar cyclers, we consider the $6\times6$
monodromy matrix $M$ of the corresponding periodic orbit in the spatial
restricted three-body problem.
It is obtained by integrating the full variational equations over one
period $T$.
\rev{Because the system is autonomous and Hamiltonian, $M$ has a double unit eigenvalue associated with the flow direction along the orbit and with the conserved Jacobi integral,  reflecting the
existence of a one-parameter family of periodic orbits. Symplecticity then ensures that the remaining multipliers occur in reciprocal pairs,}
$(\lambda_p,1/\lambda_p)$ and $(\lambda_v,1/\lambda_v)$.
We therefore define the planar and vertical stability indices,
\begin{equation}
s_p=\tfrac12(\lambda_p+1/\lambda_p),
\qquad
s_v=\tfrac12(\lambda_v+1/\lambda_v).
\label{eq:nu}
\end{equation}
A symmetric cycler is a fixed point of the area-preserving return map on the symmetry line, $\{\dot x = 0\}$ (Fig.~\ref{fig:bifurcation}b).
It is linearly stable when $|s_p|<1$ and $|s_v|<1$, so all nontrivial multipliers lie on the unit circle and the fixed point is elliptic. 
If either $|s_p|>1$ or $|s_v|>1$, the corresponding multiplier pair is real and the fixed point is hyperbolic. 
The critical values $s_p=\pm1$ and $s_v=\pm1$ mark planar and vertical bifurcations, respectively.

The planar and vertical stability mechanisms are fundamentally different. 
Because planar motion lies in an invariant submanifold of the spatial problem, the out-of-plane variation decouples from it: 
\rev{a small out-of-plane displacement $z$, a linear deviation from the planar orbit (in contrast to the nonlinear in-plane coordinates $x,y$ of Eq.~\eqref{eq:eom}),} satisfies the scalar Hill equation, 
\begin{equation}
    \ddot z+\bar U_{zz}(t)\,z=0,
\end{equation}
where $\bar U_{zz}=(1-\mu)/r_1^{3}+\mu/r_2^{3}>0$ is evaluated along the periodic orbit. 
Vertical instability can arise only through parametric resonance of this periodically varying oscillator. 
Because planar motion is invariant under the reflection
$z\mapsto -z$, the monodromy matrix is block diagonal and the
planar and vertical multiplier pairs evolve independently.
Consequently, complex instability cannot occur and $s_p$ and $s_v$
cannot exchange stability.
The in-plane dynamics, by contrast, inherit the Coriolis coupling and the hyperbolic directions associated with the $L_1$ neck from which the cyclers emerge. 
There, instability is  generic, and the primary requirement for stability is $|s_p|<1$.

Whether  vertical resonance is  encountered is an empirical question.  
Across the nine Earth–Moon families with $1\le k_1,k_2\le3$, vertical 
instability $|s_v|>1$  
occurs almost exclusively on the planar-unstable branch.
On the planar-stable branch, $|s_v|<1$ in seven of the nine families, with $\max|s_v|$ ranging from $0.57$ to $0.99$. 
In the remaining two, 
the planar-stable cap at birth is 
only marginally vertically unstable ($|s_v|\le1.04$)
and lies near the period-doubling boundary $s_v=-1$, where three-dimensional cyclers bifurcate. 
In all nine families examined, a subfamily stable in both indices exists. 
These results suggest that spatial stability is governed primarily by the planar index, while vertical instability arises only through isolated vertical bifurcations and does not preclude the existence of stable subfamilies.

\textit{Stable cyclers across mass ratio.}
Continuation in the mass ratio $\mu$
reveals stable
families across more than two orders of magnitude,
from the Sun--Jupiter regime ($\mu=10^{-3}$) to the equal-mass
limit ($\mu=\tfrac12$); representative examples are shown
in Fig.~\ref{fig:gallery} and listed in Table~\ref{tab:mu}.
The examples span several distinct cycler families, including
$(1,1)$, $(3,1)$, $(3,2)$, and $(3,3)$ families. 
The persistence of stable families across this broad range of mass
ratios supports the view that stable ballistic prograde cyclers are a
generic feature of the restricted three-body problem.

\textit{Family structure.}
The Earth--Moon system provides the clearest view of cycler family structure.
As the Jacobi constant varies, cycler families alternate between stable and unstable intervals.
\rev{The families are predominantly unstable, with stability confined to narrow intervals. For the Earth--Moon $(3,2)$ family, the two planar-stable intervals together occupy about
$0.1\%$ of its perilune-altitude range above the lunar surface
and less than $10^{-5}$ in Jacobi constant, while the median
$|s_p|$ along the family exceeds $10^8$. Nevertheless, stable
intervals can be physically appreciable: the largest $(3,3)$
interval spans about 2000~km in perilune altitude~\cite{RoRo2025Cyclers}.}
Within each stable interval, the cycler appears as an elliptic fixed point of the return map and is surrounded by invariant tori (Fig.~\ref{fig:bifurcation}b).
Additional stable intervals occur farther along the family as it approaches collision with the smaller primary, $m_2$, suggesting repeated changes of stability. 
Their detailed bifurcation structure remains to be understood.

\textit{Universality of stable subfamilies.} 
A simple argument explains why stable subfamilies should be expected.
For $C>C_1$, transport through the $L_1$ neck is impossible and no cycler exists. 
For sufficiently small $C_1-C$, the global orbit structure theorem guarantees oscillatory transit orbits associated with the stable and unstable manifolds of the $L_1$ Lyapunov orbit \cite{KoLoMaRo2000}.
A cycler family must therefore first appear at some critical value
$C_{(k_1,k_2)}^{\rm bif}
<C_{(k_1,k_2)}
<C_1$.

In the return map, the appearance of fixed points from none is generically governed by a saddle-center bifurcation~\cite{Meiss1992,MeHaOf2009,golubitsky1987generic}. 
At such a bifurcation an elliptic and a hyperbolic fixed point are created simultaneously when $\lambda_p=1$ ($s_p=1$). 
As the parameter moves away from the bifurcation, one branch becomes planar-stable ($|s_p|<1$) while the other becomes planar-unstable ($|s_p|>1$). 
Thus, saddle-center birth immediately implies the existence of a
planar-stable branch.

Every family computed exhibits precisely this structure,
spanning all values of $k_1$, $k_2$, and $\mu$ examined here
(Fig.~\ref{fig:bifurcation}).
This motivates the following conjecture: every symmetric $(k_1,k_2)$-cycler family is created at its maximal Jacobi constant through a saddle-center bifurcation of the return map. 
Combined with the observation that vertical instability is confined to isolated bifurcation intervals, this would imply that every cycler family contains a fully stable subfamily.

A proof would require showing that cycler families can arise only
through saddle-center bifurcations.
The  construction supports this picture: the image $P(\mathcal S_{k_1}^{u_1})$ intersects $U_2\cap\{\dot x=0\}$ in at most two points, so at fixed $(\mu,C)$ a symmetric family contributes at most two cyclers. 
This  restriction is consistent with saddle-center birth as the universal mechanism underlying cycler families.

\begin{table*}[!t]
\caption{
Representative fully stable symmetric $(k_1,k_2)$-cyclers spanning
more than two orders of magnitude in mass ratio, from the
Sun--Jupiter regime ($\mu=0.001$) to the equal-mass limit
($\mu=\tfrac12$).
Each orbit is specified by its perpendicular $x$-axis crossing $x_0$; 
the remaining initial conditions satisfy
$y_0=\dot x_0=0$, with $\dot y_0$ determined by the Jacobi constant
$C$. 
Here $T$ is the period and $(s_p,s_v)$ are the planar and vertical
stability indices.
All listed cyclers are fully stable, satisfying $\max\{|s_p|,|s_v|\}<1$.
}
\label{tab:mu}
\begin{ruledtabular}
\setlength{\tabcolsep}{0pt}
\begin{tabular}{lcrlrrr}
\multicolumn{1}{c}{$\mu$} & $(k_1,k_2)$ & \multicolumn{1}{c}{$x_0$} & \multicolumn{1}{c}{$C$} & \multicolumn{1}{c}{$T$} & \multicolumn{1}{c}{$s_p$} & \multicolumn{1}{c}{$s_v$} \\
\hline
$0.001$ & $(1,1)$ & -0.647047499999966 & 
 3.031605708907296 & 14.774502790974823 &  0.4121 & -0.2943  \\
$0.012150584270572$ & $(1,1)$ & -0.768217354461248 & 
 3.151175879917331 & 10.291893641936499 &  0.8210 & 0.6358  \\
$0.012150584270572$ & $(3,3)$ & -0.322477620583087 & 3.183379082910527 & 19.503763587070285 &  0.9855 & 0.6207 \\
$0.1$ & $(3,2)$ & -0.694376003123377 & 3.573367616904619  & 12.295263874014290 &  0.5686 & 0.9175 \\
$0.3$ & $(3,1)$ & -0.804725783387797 & 3.701958166478617 & 9.094576400494693 &  0.0294 & 0.8307 \\
$0.5$ & $(1,1)$ & -0.519689929077496 & 3.628400000000000 & 8.792013561462247 & 0.9376 & 0.2130 \\
\end{tabular}
\end{ruledtabular}
\end{table*}

\textit{Outlook.} 
These results identify the first continuous families of stable ballistic prograde cyclers together with the dynamical mechanism responsible for their stability. 
The persistence of stable cyclers from the Sun--Jupiter regime to the equal-mass limit suggests that they are a generic feature of binary gravitating systems rather than a peculiarity of any particular mass ratio.

These families are computed in the restricted problem, yet their periodic orbits admit continuation into the full three-body problem.
Letting the third-body mass be the small parameter $\varepsilon$, a cycler together with the primaries' circular orbit forms a periodic solution of the reduced planar flow at $\varepsilon=0$, with monodromy multipliers
$\{1,1,\lambda_p,1/\lambda_p,e^{\pm iT}\}$
consisting of the planar pair together with the synodic pair $e^{\pm iT}$ associated with the primaries’ Keplerian motion.
By the continuation theorem for periodic orbits~\cite{Hadjidemetriou1975,MeHaOf2009}, every elementary orbit, 
satisfying $\lambda_p\neq1$ and $T\not\equiv0\pmod{2\pi}$,
continues to a periodic solution of the full three-body problem for sufficiently small $\varepsilon$.
Every cycler in an open stability window is elementary\rev{, except possibly at isolated members with $T\equiv0\pmod{2\pi}$, which do not disconnect the window,} and therefore satisfies the hypotheses of the continuation theorem.
Moreover, because unit-circle multipliers vary continuously and can leave the circle only through a Krein collision~\cite{Krein1950,Moser1958}, a linearly stable planar cycler continues as a linearly stable solution except where the synodic pair $e^{\pm iT}$ resonates with the planar pair.
Finally, because the reflection symmetry $z\mapsto-z$ persists in the spatial restricted and elliptic restricted problems, the vertical multiplier pair remains decoupled and encounters no Krein obstruction~\cite{Katopodis1979}.
Thus the only caveat is an in-plane synodic resonance.

Several directions follow. 
First, the persistence of stable cyclers under small in-plane perturbations such as primary eccentricity should be investigated.
The effect of genuinely three-dimensional perturbations, including
lunar inclination and the full Earth--Moon--Sun geometry~\cite{rosengren2026}, can then be
investigated in progressively more realistic models, culminating in
full ephemeris integrations.
Second, the construction extends naturally to asymmetric, spatial, and exterior cyclers.
In the spatial problem, the vertical bifurcation points $s_v=\pm1$
generate families of three-dimensional cyclers, analogous to the
branching of halo orbits from the planar Lyapunov
family~\cite{Henon1973,BrBr1979}.

Because stable cyclers lie adjacent to chaotic transport channels, modest control inputs may enable transfers between cycler neighborhoods and distant regions of phase space~\cite{braik2026orbitalnetworksthreebodyproblem}. 
Structural stability of the saddle-center bifurcation together with persistence of the surrounding invariant tori suggests that stable cyclers are intrinsically robust dynamical structures.
More fundamentally, allowing the third body to possess non-negligible mass raises the possibility of stable ballistic planets that alternately orbit the members of a binary star system. 
Whether such cycling planets exist is an open question.

\begin{acknowledgments}
We thank Aaron J. Rosengren for helpful discussions on the historical
development of cyclers and for bringing several relevant references to
our attention.
This work was supported by the Air Force Office of Scientific Research (AFOSR)
under Grant No.~FA9550-24-1-0194.
\end{acknowledgments}

\bibliographystyle{apsrev4-2}
\bibliography{refs_cycler_pre}

\end{document}